\documentstyle[preprint,aps]{revtex}
\input{epsf}
\begin{document}
\tighten

\draft
\title{Analytical Results for Random Band Matrices with Preferential
       Basis}

\author{Klaus Frahm and Axel M\"uller--Groeling}

\address{Service de Physique de l'\'Etat condens\'e. 
	Commissariat \`a l'Energie Atomique Saclay, 
	91191 Gif-sur-Yvette, France}

\date{\today}

\maketitle

\begin{abstract}

Using the supersymmetry method we analytically calculate the local density 
of states, the localiztion length, the generalized inverse participation 
ratios, and the distribution function of eigenvector components for the 
superposition of a random band matrix with a strongly fluctuating diagonal 
matrix. In this way we extend previously known results for ordinary band 
matrices to the class of random band matrices with preferential basis. 
Our analytical results are in good agreement with (but more general than) 
recent numerical findings by Jacquod and Shepelyansky.

\end{abstract}

\narrowtext
Banded matrices with random elements play an important role in 
the description
of both certain systems exhibiting quantum chaos (in particular the kicked
rotator \cite{kicked_rotor}) and electron transport in quasi 1d geometries 
\cite{iwz}. 
In a series of papers \cite{fyodorov,mirlin},
Fyodorov and Mirlin have performed a detailed analytical investigation 
of the
properties of such matrices, making use of Efetov's supersymmetry 
technique \cite{efetov}.
In recent times particular interest in random band matrices (RBM) 
with preferential
basis (PB) (realized, e.g., by a very strongly fluctuating diagonal) has 
emerged for
at least two different reasons: First, any attempt to go beyond the quasi 
1d case for
the electron transport problem, either by employing a Fokker--Planck 
approach \cite{fokpla} or the 
(equivalent \cite{equiv}) $\sigma$ model formulation \cite{iwz,mmz}, 
necessarily leads away from the isotropic
situation and introduces a model with preferential basis. 
Second, Shepelyansky \cite{shep1} recently studied
the problem of two interacting particles in a 1d random potential by 
reducing the
Hamiltonian to a RBM with PB. He found that the two--particle localization 
length is
strongly enhanced as compared to the one--particle localization length. 
This result
was reinforced, made more precise and further investigated by subsequent work 
\cite{imry,numerics}. Very recently,
Jacquod and Shepelyansky \cite{shep2} studied numerically certain 
properties of RBM with PB,
especially the local density of states, the inverse participation ratio and 
the level
spacing statistics.

In this letter, we extend the analytical treatment of Fyodorov and Mirlin 
\cite{fyodorov,mirlin} to 
the case of RBM with PB, derive explicit formulas for 
the local density of states and the localization length, and present
expressions for the
generalized inverse participation ratios and the distribution function 
of eigenvector components in terms of previously known results for RBM. 
Our results are in good agreement with the asymptotic estimates 
given in 
\cite{shep2}
and can also explain certain numerical deviations which occur in 
\cite{shep2} 
in the limited range of accessible system parameters.

We consider the random matrix
\begin{equation}
H_{ij} =  \eta_{ij} \delta_{ij} + \zeta_{ij},
\quad\quad\quad (i,j=1,\ldots,N),
\label{eq1}
\end{equation}
where the $\eta_i \equiv \eta_{ii}$ are real random numbers with the 
distribution function
$\rho_0(\eta)$. We introduce a scale parameter $W_b$ by setting
$\langle \eta^2 \rangle \approx W_b^2$. The matrix $\zeta$ is 
either symmetric
($\zeta_{ij}=\zeta_{ji}, \beta=1$) or Hermitian 
($\zeta_{ij}=\zeta^*_{ji}, \beta=2$)
with Gaussian random variables satisfying 
$\langle |\zeta_{ij}|^2\rangle = (1+\delta_{ij}(2-\beta))A_{ij}/2$. 
$A_{ij} = a(|i-j|)$ is a function that decays on the scale of the bandwidth 
$b$ and has
typical values $1/\sqrt{b}$. In other words, $\zeta$ is exactly the RBM 
considered earlier
by Fyodorov and Mirlin \cite{fyodorov} and $\eta$ introduces a PB. For 
later use we define (as in \cite{fyodorov})
$B_0=\sum_r a(r)$ and $B_2 = \sum_r a(r) r^2/2$. The class of matrices 
introduced in 
(\ref{eq1}) contains the case considered in \cite{fyodorov} as a 
particular example:
\begin{eqnarray}
\rho_0(\eta) &=& {1\over 2W_b} \Theta(W_b-|\eta|) , \nonumber\\
a(r) &=& {2\over 3} {1\over \sqrt{1+2b}} \Theta(b-|r|).
\label{eq2}
\end{eqnarray}
The importance of the $\eta_i$ is governed by the ratio $W_b/\sqrt{b}$ of 
the spacing 
$\Delta\eta\approx W_b/b$ of those
$\eta_i$ which are coupled and the typical value $\zeta_{\mbox{\scriptsize 
typ.}} 
\approx b^{-1/2}$
of the coupling matrix elements.
We can distinguish three important regimes: (i) $W_b/\sqrt{b} \gg 1 
\Rightarrow$ the coupling matrix elements are very weak and can be treated 
perturbatively, 
(ii) $1/\sqrt{b}\ll W_b/\sqrt{b} \ll 1 \Rightarrow$ the $\eta_i$ are
still much larger than the $\zeta_{ij}$, but the coupling to the 
$\zeta_{ij}$ becomes
nontrivial, and (iii) $W_b/\sqrt{b} \lesssim 1/\sqrt{b} \Rightarrow$ the 
$\eta_i$ are comparable
to the $\zeta_{ij}$ and the RBM results in \cite{fyodorov,mirlin} are 
applicable. 

In the regime (i), one may apply simple perturbation theory to calculate 
the eigenfunctions. 
This yields the behavior $|\psi(j\,b)|\sim 
|\psi(0)|\,(\zeta_{\mbox{\scriptsize typ.}}/
\Delta \eta)^j$. The corresponding localization length is then 
estimated as $\xi\sim 2b/\ln(W_b^2/b)$, a behavior that was also 
anticipated 
(and numerically confirmed) in \cite{shep1}. In the present work, we are 
mostly concerned with regime (ii) and the crossover to (iii). 

We are interested in calculating the (position dependent) generalized 
inverse participation
ratios (we use the notation of \cite{mirlin}):
\begin{eqnarray}
P_q(E,n) &=& {1\over \rho(E)} \left\langle
\sum_k \left\vert \psi_k(n) \right\vert^{2q} \delta(E-E_k) 
\right\rangle \nonumber\\
&=& \lim_{\varepsilon \to 0}
{i^{l-m} \over 2\pi\rho} (2\varepsilon)^{q-1}
{(l-1)!(m-1)! \over (l+m-2)!}
\left\langle (G^+_{nn})^l (G^-_{nn})^m \right\rangle
\quad\quad\quad (q=l+m).
\label{eq3}
\end{eqnarray}
To perform the average over the product of Green's functions we use the 
supersymmetry method.
For all details and the standard notation we have to refer the reader 
to \cite{efetov,vwz}.

We define a supersymmetric functional
\begin{eqnarray}
F(J) &=& \left\langle \mbox{sdet}^{-1/2}(E-H+i\varepsilon\Lambda + 
\hat{J}) 
         \right\rangle_\zeta = \left\langle 
\mbox{sdet}^{-1/2}(1+g\hat{J} \right\rangle_\zeta
         \nonumber \\
  &=& \left\langle 
(1+x_+G_{nn}^+)^{-1}(1+x_-G_{nn}^-)^{-1} \right\rangle_\zeta,
\label{eq4}
\end{eqnarray}
where $g=\mbox{diag}(G^+,G^-)$, and $\hat{J} = \mbox{diag}(x_+P_B,x_-P_B) 
\otimes 
e_ne_n^\dagger = J\otimes e_ne_n^\dagger$. Here, $x_+$ and $x_-$ are simple 
numbers, $P_B$ is
a projector on bosonic variables, $e_n$ is a vector with $(e_n)_i 
= \delta_{ni}$,
and $\langle\ldots\rangle_\zeta$ denotes averaging over $\zeta$. Our 
formulas are valid
for both orthogonal and unitary symmetry, provided the four--dimensional 
graded space in the
unitary case is simply doubled. Following standard procedures 
\cite{fyodorov,mirlin}, 
we express (\ref{eq4})
as an integral over supervectors, average over $\zeta_{ij}$, perform a 
Hubbard--Stratonovitch transformation and finally arrive at
\begin{eqnarray}
F(J) &=& \int D[\sigma_i] \exp\{ - {\cal L}_1(\sigma) \}, \nonumber\\
{\cal L}_1(\sigma) &=& {1\over 8} \sum_{ij} \mbox{str} (\sigma_i 
(A^{-1})_{ij} \sigma_j)
+{1\over 2} \sum_j \mbox{str}\ln(E-\eta_j+i\varepsilon\Lambda +\sigma_j/2 
+ J\delta_{jn}),
\label{eq5}
\end{eqnarray}
where the $\sigma_i$ are $8\times 8$ supermatrices. Note that the 
functional still
depends on the particular realization of the $\eta_j$. No average over 
these variables
has been performed yet. The saddle--point condition for (\ref{eq5}) reads
\begin{equation}
\sigma_j = -\sum_l A_{jl} {1\over 
E-\eta_l+i\varepsilon\Lambda +\sigma_l/2}.
\label{eq6}
\end{equation}
With the ansatz $\sigma = \Gamma_0 +
i\Gamma_1 Q$ for the solution of (\ref{eq6}), where $Q$ satisfies $Q^2=1$ 
and is a proper
element of the coset space introduced in \cite{efetov}, we have 
for $z=\Gamma_0+i\Gamma_1$:
\begin{equation}
z = -B_0\int d\eta \rho_0(\eta) {1\over E-\eta+z/2}
\quad\quad\quad (\mbox{Im}(z)>0).
\label{eq7}
\end{equation}
Here we have replaced each element of the sum in (\ref{eq6}) by its 
average over
$\eta$. This procedure is only justified in the limit of 
``overlapping resonances'',
i.e. if the widths $\Gamma_1$ of the superimposed Lorentzians are larger 
than their
spacings determined by those $\eta_l$ contributing to the sum. With 
the result
$\Gamma_1 \sim \rho(E) \sim 1/W_b$ (see (\ref{eq16}) below) this 
immediately leads back
to the condition $W_b \ll \sqrt{b}$, which is fulfilled in our regimes (ii) 
and (iii)
defined above. Also, in precisely this limit we can expect the 
saddle--point solving
(\ref{eq6}) to be homogeneous in space as in our ansatz.

Equation (\ref{eq7}) determines $\Gamma_0$ and $\Gamma_1$ and hence the 
local density of
states $\langle \rho(j,E) \rangle_\zeta$. Using (\ref{eq4}) it is not 
difficult to show that
$\langle G_{jj}^+\rangle_\zeta = (E-\eta_j+z/2)^{-1}$ and therefore
\begin{equation}
\langle \rho(j,E)\rangle_\zeta = -{1\over\pi} Im 
\langle G_{jj}^+(E)\rangle_\zeta
= {\Gamma_1/2\pi \over (E-\eta_j+\Gamma_0/2)^2 + \Gamma_1^2/4}.
\label{eq15}
\end{equation}
This is exactly the Breit--Wigner form found numerically in \cite{shep2}. 
Averaging this expression
over all sites (assuming that $N^{-1}\sum_j(\ldots) = \langle \ldots 
\rangle_\eta$) and using
the saddle--point equation (\ref{eq7}) we get
\begin{equation}
\rho(E) \equiv \langle\langle \rho(j,E)\rangle_\zeta\rangle_\eta = 
\Gamma_1/\pi B_0
\label{eq16}
\end{equation}
for the average density of states. From (\ref{eq7}) and (\ref{eq16}), we 
find the 
limiting cases $\rho(E)\simeq \rho_0(E)$ for $W_b\gg 1$ and $\rho(E)\simeq 
\sqrt{2B_0-E^2}/(\pi B_0)$ for $W_b\ll 1$.

Inserting the saddle--point solution in (\ref{eq5}) we proceed in analogy 
to 
\cite{fyodorov,mirlin} and arrive at
\begin{eqnarray}
F(J) &=& \int D[Q_i] \exp\{-{\cal L}_2(Q)\}, \nonumber\\
{\cal L}_2(Q) &=& -{\xi\over 16} \sum_j \mbox{str}[(Q_{j+1}-Q_j)^2] + 
{1\over 2} \sum_j \mbox{str}\ln 
[1+M_j(Q_j)(i\varepsilon\Lambda+J\delta_{jn})], \nonumber\\
M_j(Q) &=& (E-\eta_j+\Gamma_0/2 + i\Gamma_1/2 Q)^{-1}.
\label{eq8}
\end{eqnarray}
Here, $\xi = 2\Gamma_1^2 B_2/B_0^2 = 2\pi^2 \rho(E)^2 B_2$ denotes the 
localization 
length \cite{larkin,efetov} for the wave functions \cite{lokdiscuss}, 
$|\psi(n)|\sim|\psi(0)|\exp(-n/\xi)$. 
Due to the one--dimensional structure
of the $Q$--functional (\ref{eq8}) $F(J)$ can be written as
\begin{equation}
F(J) = \int dQ\, Y(Q,n)\, Y(Q,N-n)\, \mbox{sdet}^{-1/2}(1+M_n(Q)J),
\label{eq9}
\end{equation}
where the function $Y(Q,n)$ arises from integrating out all $Q$--matrices 
up to site $n$. The
properties of these functions have been discussed in detail in 
\cite{fyodorov,mirlin}. From (\ref{eq4}) it is
clear that
\begin{equation}
\langle (G_{nn}^+)^l (G_{nn}^-)^m \rangle = 
{(-1)^{l+m} \over l!m!} \partial^l_{x_+} \partial^m_{x_-}
 F(J)  \Big\vert_{x_\pm = 0}.
\label{eq9a}
\end{equation}

Performing the source term derivatives in (\ref{eq9}) and using that
\begin{equation}
M_j(Q) = {E-\eta_j+\Gamma_0/2 - i(\Gamma_1/2) Q \over
          (E-\eta_j+\Gamma_0)^2 + \Gamma_1^2/4 }
\label{eq10}
\end{equation}
we finally get in analogy to \cite{mirlin} (for $\varepsilon\to 0$
and $\beta=2$)
\begin{equation}
\langle (G_{nn}^+)^l (G_{nn}^-)^m \rangle = 
{l+m \choose m}\left(-i\pi \rho\,C(\eta_n)\right)^{l+m}\,
\int dQ\,Y(Q,n)\,Y(Q,N-n)\,Q_{11,BB}^l\,Q_{22,BB}^m
\label{eq11}
\end{equation}
with
\begin{equation}
C(\eta_n) = {1\over \pi\rho}
{\Gamma_1/2 \over (E-\eta_n+\Gamma_0/2)^2 + \Gamma_1^2/4 }.
\label{eq13}
\end{equation}
Comparing (\ref{eq11}) with the corresponding expressions in \cite{mirlin} 
we can immediately conclude that
\begin{equation}
P_q(E,n) = C(\eta_n)^q P_q^{FM}(E,n).
\label{eq12}
\end{equation}
The superscript ``FM'' refers to the result by Mirlin and Fyodorov 
\cite{mirlin} 
for ordinary RBM, where their rescaled length $x$ has to be identified 
with $x=2N/(\beta\xi)$ [$\xi$ as in eq. (\ref{eq8})]. 
Similarly, following the steps in \cite{mirlin} to derive the {\em local} 
distribution function ${\cal P}(y,n)$ for $y=N|\psi(n)|^2$
defined by
$P_q(E,n) = N^{1-q} \int_0^\infty dy \,{\cal P}(y,n)\, y^q$, we find 
\cite{distdiscuss} 
\begin{equation}
{\cal P}(y,n) = {1\over C(\eta_n)} 
{\cal P}^{FM}\left(\frac{y}{C(\eta_n)},\,n\right).
\label{eq14}
\end{equation}
We see that the local distribution is rescaled by the energy dependent 
factor 
$C(\eta_n)$. In the resonant case, $E\simeq \eta_n$, the typical amplitude 
$y$ 
exhibits 
a strong enhancement (in the limit $W_b\gg 1$, i.e. $\Gamma_1\sim W_b^{-1}$),
whereas for $(E-\eta_n)^2\gg B_0$ a strong suppression is observed. 

Up to now we have considered a fixed realization of the diagonal elements 
$\eta_j$. 
The additional average over the sites (or equivalently over $\eta_n$) 
results 
in
\begin{equation}
\label{eq14a}
{\cal P}(y)=\int d\eta\ \rho_0(\eta)\,
\frac{1}{C(\eta)}\,{\cal P}^{FM}\left(\frac{y}{C(\eta)}\right).
\end{equation}
This completes our formal derivations for the case of RBM with PB. In the 
following we 
derive some more explicit results for the special case of Shepelyansky 
(see (\ref{eq2})).

With $B_0=2/3$ we have $\rho(E) = 3\Gamma_1(E)/2\pi$, where $\Gamma_1(E)$ 
is determined
by (\ref{eq7}) and depends on $E$ and $\rho_0(\eta)$ (i.e. on $W_b$ for the 
case (\ref{eq2})).
In the main part of Fig.1 we show $\rho(E)$ for different values of 
$W_b$, demonstrating
the crossover from a semicircle to the box shape (\ref{eq2}).
For $E=0$ we must have $\Gamma_0=0$ and (\ref{eq7}) simplifies to
\begin{equation}
\Gamma_1(0) = {2\over 3W_b} \arctan \left( {2W_b\over \Gamma_1(0)} 
\right) .
\label{eq17}
\end{equation}
With $B_2=b^2/9$ we have $\xi = \Gamma_1(0)^2b^2/2$ so that we can express 
$\xi$ as a function
of $W_b$. We recall that our treatment is valid for all $W_b$ with
$0 \le W_b \ll \sqrt{b}$, i.e. for RBM, RBM with PB and the whole 
crossover. In the inset of
Fig.1 we have plotted $\xi$ versus $W_b$. Interestingly, the crossover 
region between the two
limiting cases $\xi/b^2= const.$ (RBM) and $\xi/b^2 \sim 1/W_b^2$ (RBM with 
PB as specified in
(\ref{eq2})) is a comparatively small interval around $W_b=1$.

For the asymptotic case $W_b\gg 1$ considered in the estimates in 
\cite{shep1,shep2} we get by
expanding (\ref{eq17})
\begin{equation}
\xi \approx {\pi^2\over 18} {b^2\over W_b^2} (1-{2\over 3W_b^2}).
\label{eq18}
\end{equation}
This is in very good agreement with the estimate $b^2/2W_b^2$ in 
\cite{shep1,shep2}. 
We also remark
that a naive application of the RBM results in \cite{fyodorov} (where one 
simply modifies the function
$a(|r|)$ to account for the large diagonals) does reproduce the parameter 
dependence in 
(\ref{eq18}) but yields a (wrong) prefactor $4/3$ instead of $\pi^2/18$.

Finally, we calculate the quantity $\xi_{IPR}$, defined by 
$\xi_{IPR}^{-1} = P_2 = N^{-1}\sum_nP_2(E,n)$ in the limit $W_b\gg 1$.
We get for arbitrary $\xi$ and $N$
\begin{equation}
\xi_{IPR} = {\pi^2\over 12W_b^2} \left( {3\over 2N} + {1\over 
\xi} \right)^{-1},
\label{eq19}
\end{equation}
with the metallic limit ($N\ll \xi$)
\begin{equation}
\xi_{IPR} = {\pi^2\over 18} {N\over W_b^2} \approx 0.548 {N\over W_b^2}
\label{eq20}
\end{equation}
and the localized limit ($N\gg \xi$)
\begin{equation}
\xi_{IPR} = {\pi^2 \over 12W_b^2} \xi = {\pi^4 \over 6^3} {b^2 \over 
W_b^4} \approx
0.451 {b^2\over W_b^4}.
\label{eq21}
\end{equation}
These analytical results compare quite favourably with the estimates \
$N/2W_b^2$ (resp. $b^2/4W_b^4$) in \cite{shep2}. Also, in view of the exact
formula (\ref{eq19}), which interpolates between the metallic and 
the localized
regime, the numerical deviations from $\xi_{IPR} \sim W_b^{-2}$ 
(resp. $W_b^{-4}$)
found in \cite{shep2} for non--asymptotic system parameters come as 
no surprise.

Concerning the distribution (\ref{eq14a}), let us briefly discuss the 
simplest case (with 
$\beta=2$ and $\xi\gg N$), where ${\cal P}^{FM}(y)=\exp(-y)$ as 
for a GUE random matrix. 
At $E=0$ and $W_b\gg 1$ [$\Gamma_1=\pi/(3W_b)$], we find:
\begin{equation}
\label{eq22}
{\cal P}(y)=\frac{1}{W_b}\int^{W_b}_0 d\eta\,3(\eta^2+\Gamma_1^2/4)
\,\exp[-3(\eta^2+\Gamma_1^2/4)\,y].
\end{equation}
The probability is shifted to very 
small amplitudes with $y\ll W_b^{-2}$, i.e. ${\cal P}(y)\sim W_b^2$,
and also to
higher amplitudes $y\gg W_b^{-2}$,  
${\cal P}(y) \sim \exp[-\pi^2 y/(36 W_b^2)]$. 

In conclusion we have applied the supersymmetry method to the class of RBM 
with PB. The
key to technical progress was the derivation of a $\sigma$ model for a 
fixed realization
of the strongly fluctuating diagonal matrix elements. The average over 
these diagonal 
variables was then performed in a second step, whenever appropriate. Our 
results account
for the complete crossover between ordinary RBM (as treated earlier in 
\cite{fyodorov,mirlin}) 
and RBM with
strong PB. For the specific model chosen in 
recent numerical calculations \cite{shep2} the agreement
with our findings is very satisfactory. In particular, the existence
of two different scales characterizing the wavefunction as evidenced
by the difference between $\xi_{IPR}$ and $\xi$ in (\ref{eq19}) has been
demonstrated analytically.
We believe that further investigations of RBM with
PB along the lines given in this letter are interesting and feasible.

We are grateful to J.--L. Pichard and D. Shepelyansky for instructive
discussions.
This work was supported by the European HCM program (KF) and a 
NATO fellowship
through the DAAD (AMG).
Upon completion of this manuscript a preprint with very similar results by 
Y.V.Fyodorov and
A.D.Mirlin (cond-mat/9507043) came to our attention.


\begin{figure}
\epsfxsize=6in
\epsfysize=4in
\epsffile{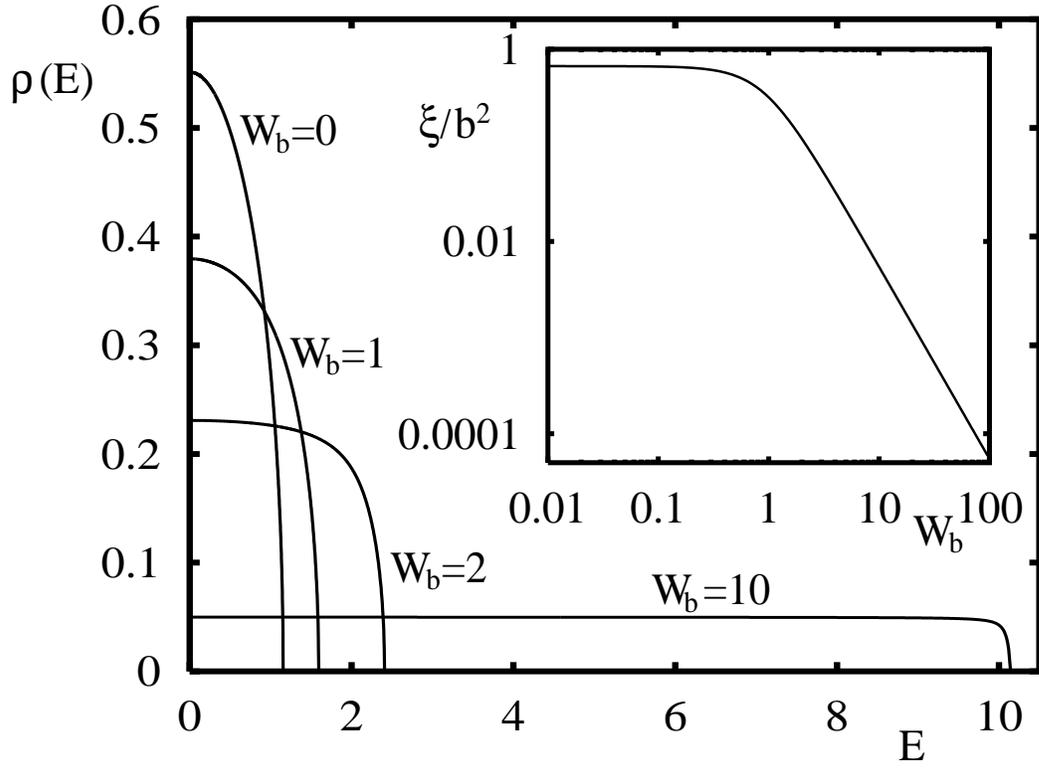}
\smallskip
\caption{The average density of states for the particular band matrix 
considered 
in Refs. \protect\cite{shep1,shep2} (see Eq. (\protect\ref{eq2})) and the 
values 
$W_b=0,\ 1,\ 2,\ 10$. The inset shows the localization length $\xi$ 
normalized 
by $b^2$ at $E=0$ as a function of $W_b$ (in a doubly 
logarithmic representation).}
\label{fig1}
\end{figure}


\begin{thebibliography}{99}
\bibitem{lokdiscuss} We have to consider that $\langle |G_{1n}|^2\rangle 
\sim \exp[-n/(2\xi)]$ when applying the established $\sigma$ model results 
of \cite{larkin,efetov} to determine the localization length. Note that 
$|G_{1,n}|^2_{\mbox{\scriptsize typical}}\sim |\psi(n)|^2\sim 
\exp(-2n/\xi)$ 
due to the log--normal distribution of $|G_{1,n}|^2$.
\bibitem{distdiscuss} In Ref. \cite{mirlin}, the site averaged 
distribution 
function ${\cal P}^{FM}(y)=N^{-1} \sum_n {\cal P}^{FM}(y,n)$ was 
calculated. 
The ``local'' distribution ${\cal P}^{FM}(y,n)$ depends only weakly on the 
site $n$, as long as we deal with the metallic regime ($\xi\gg N$) or with 
the 
localized regime ($\xi\ll N$) far from the boundary ($n, N-n\gg \xi$). In 
Eq. (\ref{eq14}) the main $n$--dependence arises from the rescaling factor 
$C(\eta_n)$.
\end{thebibliography}
\end{document}